\documentclass[11pt]{article}
\usepackage{floatflt}
\usepackage[dvips]{graphics,epsfig}
\begin{document}
\def\thesection{\arabic{section}}
\def\thesubsection{\arabic{section}.\arabic{subsection}}
\def\ka{\hbox{\ae}}
\def\kappa{\hbox{\ae}}
\setcounter{page}{927} \noindent {\small Bulletin of the Russian Academy of Sciences
\hfill Vol. 60, No 6, pp. 927-945}

\noindent {\small Physics \hfill 1996}\\
\copyright 1996 {by Allerton Press, Inc.\vspace{10mm}%\hrule\vspace{20mm}

\noindent {\large \bf INVERSIONLESS AMPLIFICATION AND LASER-INDUCED\\ TRANSPARENCY AT
THE DISCRETE TRANSITIONS\\  AND THE TRANSITIONS TO CONTINUUM}
\footnote{Invited review paper}\\

\noindent {\large A. K. Popov\\

\noindent{\large \it Kirensky Physics Institute, Siberian Division of the  Russian
Academy of Sciences, \\ Krasnoyarsk 660036}\,\
e-mail: popov@ksc.krasn.ru}\\

%\title{ INVERSIONLESS AMPLIFICATION AND LASER-INDUCED\\
%TRANSPARENCY AT THE DISCRETE TRANSITIONS \\
%AND THE TRANSITIONS TO CONTINUUM}
%\author{A.K. Popov\\
%Kirensky Physics Institute, Siberian Division of the\\ Russian
%Academy of Sciences, Krasnoyarsk 660036}
%\date{}
%\hoffset -25mm \voffset -15mm \topmargin=0mm \textheight=250mm
%\textwidth=170mm \columnsep=8mm
%%\textheight = 24 cm
%%\textwidth = 16 cm
%\begin{document}
%\def\thesection{\arabic{section}}
%\def\thesubsection{\arabic{section}.\arabic{subsection}}
%\setcounter{page}{927}
%\begin{abstract}
The effects of coherence of quantum transitions and the interference of resonant
nonlinear optical processes on the spectra of absorption, amplification, and
nonlinear-optical generation are considered. The most favorable conditions are
discussed for the inversionless amplification, resonant refraction in the absence of
absorption and for resonant enhancement of nonlinear-optical generation  at the
discrete transitions and the transitions to continuum.\\

%\end{abstract}

\section{Introduction}
Coherency and interference are the basic physical phenomena, which lead to new
effects in quantum optics. Interference can be destructive or, vice versa,
constructive, causing mutual suppression or amplification of simultaneous processes.
In quantum optics the interference effects can reduce interaction of radiation with
absorbing atoms at lower energy levels without substantial variations in the
interaction with emitting atoms at overlying levels. In turn, this leads to a
difference in absorption and emission spectra. The intratomic coherence conditions
many fundamental effects in high-resolution nonlinear spectroscopy, the light
amplification without population inversion and resonant refraction increase in the
absence of absorption, coherent population trapping, increase of the resonant
nonlinear-optical radiation at a simultaneous primary radiation absorption decrease,
and the laser induction of structures of the autoionization type in spectral
continua. The resonant nonlinear interference effects, theoretically and
experimentally studied since the time of first masers and lasers creation [1-6],
currently again attracted a great attention [7-10]. They are promising for new laser
sources in the VUV and X-ray bands, laser accelerators of atomic particles,
microscopes with increased resolution, supersensitive magnetometers, etc. (see, e.g.
[11]. A lot of scientific meetings are dedicated to this problem (see, e.g. [9, 10]).

A publication flux are devoted to the related effect of electromagnetically induced
transparency as applied to improve characteristics of laser light conversion to
short-wave bands (see. e.g. [7]). (It appears that in many respects this is the
phenomenon comprehensively studied in Russia in 1960-70~s [12-36, 40]). A quite
complete survey of results acquired by western authors can be found in [7-10] and
references therein. Therefore, below a basic attention would be given to certain
less-known results of Russian authors.

A basic contribution into studies of the nonlinear interference processes in
absorption (emission) spectra at the interaction of atom-molecular systems with
electromagnetic radiation and the effect of dynamic splitting of spectral lines in
strong electromagnetic fields was made by the Yerevan, Moscow, Nizhny Novgorod,
Novosibirsk, St. Petersburg, and Minsk schools. These results are generalized in
monographs [4, 15-18, 22, 23, 30, 33, 36]. In [25, 28, 29, 30] the possible
inversionless amplification was analyzed for three-level systems at the discrete
optical transitions. Corresponding experimental studies were carried out in [21, 33,
34]. (For two-level optical systems this effect was predicted in [5] and
experimentally studied in the radio band in [6, 13, 14]). The effect of
self-transparency induced by a strong field at an adjacent transition was
theoretically and experimentally studied in detail, e.g. in [12, 19, 20]. The
coherent population trapping was first observed in [35]. Later the study in nonlinear
interference phenomena at the discrete transitions was extended to the transitions to
a continuous spectrum [31, 32, 36], autoionization-type resonances were predicted in
[31] and experimentally revealed [32] at the atomic transitions to continuum. (Later
this effect was called as the laser-induced continuum structure (LICS)). They also
predicted the possible nonlinear response increase at the laser short-wave generation
with a simultaneous decrease in its absorption and improvement in phase matching [31,
36]. In [37] it was shown that, similar to discrete transitions, the inversionless
amplification is possible also at the transitions to autoionization and
antoionization-type states. The cited papers initiated the coherence effect study,
first at the transitions to continuum, and then at the discrete optical ones.

The coherence and
interference phenomena are the basis for inversionless amplification,
coherent population trapping, and electromagnetically induced transparency
both at the discrete transitions and those to continuous spectrum. As it
was already indicated, these effects offer unconventional solution for
actual problems of quantum electronics. However, the peculiarities of
optical transitions and real experimental designs can qualitatively change
an expected manifestation of these processes. These problems remain to be a
subject of great attention.
Hence, it is required to develop theoretical approaches considering the
most important accompanying processes and involving numerical analysis, if
necessary. Now the least understood phenomena are the effects of nonuniform
broadening and level degeneration, relaxation and motion of population on
the coherence degradation. It appears that sometimes, vice versa, the
relaxation promotes the intratomic coherence. A fairly small number of
papers is dedicated to the effect of above processes on the resonant
nonlinear-optical frequency mixing.

\section{ Resonant nonlinear-optical interference}
\subsection{ Destructive and constructive interference in classical and
quantum optics}

The interference is one of fundamental physical phenomena. Oscillations of various
nature depending on a phase relationship can interfere constructively or
destructively. Varying oscillation phases and amplitudes, the resulting process can
be amplified or suppressed. The quantum interference can proceed, when there is
coherent superposition of real states. Moreover, the degenerate (in frequency)
interfering intra-atomic oscillations can be conditioned by different correlating
quantum transitions contributing into the same process. These are, e.g., one-and
two-photon contributions into the optical process related to emission or absorption
at a specified frequency. The process can result from the coherent superposition of a
neighboring real energy level and a quasi-level (virtual state) created by a strong
auxiliary field [26]. Such a superposition is realized even more simply than in the
case of real doublet state.

The interference is more general concept, than notions of one-, two-, and multi-step
and multi-photon processes. The latter were introduced and classified by their
frequency-correlation properties in the framework of perturbation theory. However,
these properties are significantly varied as the field intensity rises, in particular
at resonant interaction [26]. As a result, the qualitative effects become possible in
nonlinear spectroscopy of the Doppler-broadened transitions, such as the induced
compensation of residual Doppler broadening in two-photon absorption or Raman
scattering under conditions  of a difference in photon frequencies [38, 39]. Even
when many elementary processes contribute into the optical one at a given frequency
and their classification into stepped and multi-photon processes is difficult,
experimental data can be often explained and predicted using the concept on
interfering components of nonlinear polarization. Amplitudes and phases of these
components are varied by controlling the corresponding field intensities and the
detunings from one- and two-photon resonances.

\subsection{ Equation for the density matrix: the effects of intermediate
level population and relaxation}
In the general case of an open configuration of
energy levels, when a lower level is not ground, various relaxation processes' rates
are different, and all the levels can be populated, the density matrix approach is
most convenient to analyze resonant nonlinear-optical processes. Simple formulas for
spectral properties of responses at a weak probing field frequency in the presence of
strong one at an adjacent transition are uniformly deduced for $V$, $\Lambda$, and
cascade schemes [22, 23, 27, 30, 36]. Let us show this by the example of transition
diagram displayed in Fig. 1.
\begin{floatingfigure}[l] {50mm}
\includegraphics[width=40mm]{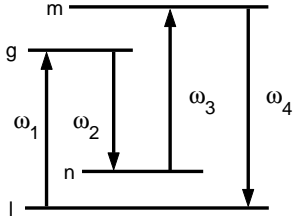}
\caption{Transition configuration.}
\end{floatingfigure}
The scheme is assumed open, i.e. level $l$ is not ground one. For simplicity, we
consider the fields $E_2$ and $E_4$ at frequencies $\omega_2$ and $\omega_4$, are
probing, i.e., they do not disturb the level population. The fields $E_1$ and $E_3$ at
frequencies $\omega_1\approx\omega_{gl}$ and $\omega_3\approx\omega_{mn}$, are strong.
Below, in Sec. 3, we eliminate this limitation while considering interaction of two
strong fields. Let us acquire the conditions of inversionless amplification at the
transitions $gn$ and $ml$ so that to consider both the configurations $V$ and
$\Lambda$. The probing field frequencies can be either higher or lower than those of
strong fields. In the interaction representation the density matrix components and
corresponding equations take the form $$ \rho_{lg}=r_1\exp(i\Omega_1 t), \quad
\rho_{nm}=r_3\exp(i\Omega_3 t),\quad \rho_{ng}=r_2\exp(i\Omega_1 t)+\tilde r_2
\exp[i(\Omega_1+\Omega_3-\Omega_4)t], $$ $$ \rho_{lm}=r_4\exp(i\Omega_4 t)+\tilde r_4
\exp[i(\Omega_1-\Omega_2+\Omega_3)t], \quad
\rho_{ln}=r_{12}\exp[i(\Omega_1-\Omega_2)t]+r_{43}\exp[i(\Omega_4-\Omega_3)t], \quad
\rho_{ii}=r_i, $$ $$ P_2r_2=iG_2\Delta r_2- iG_3r_{32}^*+ir_{12}^*G_1, \quad d_2
\tilde r_2=-iG_3 r_{41}^*+ir_{43}^*G_1, \quad P_4r_4=i[G_4\Delta r_4-G_1
r_{41}+r_{43}G_3], $$ $$ d_4 \tilde r_4=-iG_1 r_{32}+ir_{12}G_3, \quad P_{41}
r_{41}=-iG_1^* r_4+ir_1^*G_4, \quad P_{43} r_{43}=-iG_4 r_4^*+ ir_4 G_3^*, $$ $$
P_{32} r_{32}=-iG_2^* r_3 + ir_2^*G_3, \quad P_{12} r_{12}=-iG_1 r_2^*+ ir_1 G_2^*,
\quad \Gamma_m r_m=-2 {\rm Re}\{ iG_3^* r_3  \}+q_m, $$ $$ \Gamma_n r_n=-2 {\rm Re}\{
iG_3^* r_3  \}+\gamma_{gn}r_g+\gamma_{mn} r_m +q_n, \ \Gamma_g r_g=-2 {\rm Re}\{
iG_1^* r_1  \}+q_g,$$  $$\Gamma_l r_l=-2 {\rm Re}\{ iG_1^* r_1
\}+\gamma_{gl}r_l+\gamma_{ml} r_m +q_l,$$ where $ \Delta r_1=r_l-r_g, \ \Delta
r_2=r_n-r_g, \ \Delta r_3=r_n-r_m, \ \Delta r_4=r_l-r_m;$ $
\Omega_1=\omega_1-\omega_{lg}, \ \Omega_3=\omega_3-\omega_{mn}, \
\Omega_2=\omega_2-\omega_{gn}, \ \Omega_4=\omega_4-\omega_{ml}; $ $ G_1=- {E_1
d_{lg}}/{2 \hbar}, \ G_2=- {E_2 d_{gn}}/{2 \hbar}, \ G_3=- {E_3 d_{nm}}/{2 \hbar}, \
G_4= - {E_4 d_{ml}}/{2 \hbar}; $ $ P_1=\Gamma_{lg}+i\Omega_1, \
P_2=\Gamma_{ng}+i\Omega_2, \ P_3=\Gamma_{nm}+i\Omega_3, \ P_4=\Gamma_{lm}+i\Omega_4,
 P_{12}=\Gamma_{ln}+i(\Omega_1-\Omega_2), \
P_{43}=\Gamma_{ln}+i(\Omega_4-\Omega_3), \ P_{32}=\Gamma_{gm}+i(\Omega_3-\Omega_2), \
P_{41}=\Gamma_{gm}+i(\Omega_4-\Omega_1), \
d_2=\Gamma_{ng}+i(\Omega_1+\Omega_3-\Omega_4), \
d_4=\Gamma_{lm}+i(\Omega_1-\Omega_2+\Omega_3).$ Here $\Omega_i$ are the frequency
detunings from resonances, $G_i$ are the Rabi frequencies, $\Delta r_i$, are the
population differences depending on intensity, $\Gamma_{ij}$ are the uniform
halfwidths of transitions, $\Gamma_i^{-1}$ are the lifetimes, $\gamma_{ij}$ is the
rate of relaxation from the $i$th to $j$th levels, and $q_i$ are the rates of
excitation by additional noncoherent pumping. The off-diagonal density matrix
amplitudes $r_i$, define the coefficients of absorption (amplification) and refraction
indices and $\tilde r_i$, define the nonlinear polarization of four-wave mixing.  For
the cascade configurations, the equations and their solutions are deduced by a simple
substitution of detuning signs or by a complex conjugation of corresponding
co-factors.
\subsection{ Laser-induced intra-atomic coherence and classification of
resonant nonlinear-optical effects} Solution to the set of coupled equations for the
density matrix components is given by $$ r_{1,3}=i {G_{1,3}\Delta r_1}/{P_1}, \ \
r_{2,4}=i {G_{2,4}}R_{2,4}/{P_{2,4}}, $$ $$R_2=\frac { \Delta
r_2(1+g_7+v_7)-v_3(1+v_7-g_8)\Delta r_3-g_3(1+g_7-v_8)\Delta r_1}
{(1+g_2+v_2)+[g_7+g_2(g_7-v_8)+v_7+v_2(v_7-g_8)]},\eqno(1)$$ $$R_4=\frac { \Delta
r_4(1+v_5+g_5)-g_1(1+g_5-v_6)\Delta r_1-v_1(1+v_5-g_6)\Delta r_3}
{(1+g_4+v_4)+[v_5+v_4(v_5-g_6)+g_5+g_4(g_5-v_6)]}, \eqno(2)$$ $$\Delta
r_1=\frac{(1+\ka_3)\Delta n_1+b_1\ka_3 \Delta n_3}
{(1+\ka_1)(1+\ka_3)-a_1\ka_1b_1\ka_3},\ \Delta r_3=\frac{(1+\ka_1)\Delta n_3+a_1\ka_1
\Delta n_1}{(1+\ka_1)(1+\ka_3)-a_1\ka_1b_1\ka_3},$$ $$\Delta r_2=\Delta
n_2-b_2\ka_3\Delta r_3-a_2\ka_1\Delta r_1,\ \Delta r_4=\Delta n_4-a_3\ka_1\Delta
r_1-b_3\ka_3\Delta r_3;$$ $$r_m=n_m+(1-b_2)\ka_3\Delta r_3,\; r_g=n_g
+(1-a_3)\ka_1\Delta r_1,\; r_n=n_n-b_2\ka_3\Delta r_3+a_1\ka_1\Delta r_1,\eqno(3)\;$$
$$r_l=n_l-b_1\ka_3\Delta r_3+a_3\ka_1\Delta r_1,\ \Delta r_i(E_1=0, E_3=0)= \Delta
n_i;$$ $$ g_1=\frac{|G_1|^2}{P_{41}P_1^*}, g_2=\frac{|G_1|^2}{P_{12}^*P_2},
g_3=\frac{|G_1|^2}{P_{12}^*P_1^*}, g_4=\frac{|G_1|^2}{P_{41}P_4},
g_5=\frac{|G_1|^2}{P_{43}d_2^*}, g_6=\frac{|G_1|^2}{P_{41}d_2^*},
g_7=\frac{|G_1|^2}{P_{32}^*d_4^*}, g_8=\frac{|G_1|^2}{P_{12}^*d_4^*},$$
$$v_1=\frac{|G_3|^2}{P_{43}P_3^*}, v_2=\frac{|G_3|^2}{P_{32}^*P_2},
v_3=\frac{|G_3|^2}{P_{32}^*P_3^*}, v_4=\frac{|G_3|^2}{P_{43}P_4},
v_5=\frac{|G_3|^2}{P_{41}d_2^*}, v_6=\frac{|G_3|^2}{P_{43}d_2^*},
v_7=\frac{|G_3|^2}{P_{12}^*d_4^*}, v_8=\frac{|G_3|^2}{P_{32}^*d_4^*}; $$
$$\ka_1=\ka_1^0\frac{\Gamma_{lg}^2}{|P_1|^2},
\ka_1^0=\frac{2(\Gamma_l+\Gamma_g-\gamma_{gl})} {\Gamma_l\Gamma_g\Gamma_{lg}}|G_1|^2,
\ka_3=\ka_3^0\frac{\Gamma_{mn}^2}{|P_3|^2},
\ka_3^0=\frac{2(\Gamma_m+\Gamma_n-\gamma_{mn})}
{\Gamma_m\Gamma_n\Gamma_{mn}}|G_3|^2;$$
$$a_1=\frac{\gamma_{gn}a_2}{\Gamma_n-\gamma_{gn}}=
\frac{\gamma_{gn}\Gamma_la_3}{\Gamma_n(\Gamma_g-\gamma_{gl})}=
\frac{\gamma_{gn}\Gamma_l}{\Gamma_n(\Gamma_l+\Gamma_g-\gamma_{gl})}, $$
$$b_1=\frac{\gamma_{ml}\Gamma_nb_2}{\Gamma_l(\Gamma_m-\gamma_{mn})}
=\frac{\gamma_{ml}b_3}{\Gamma_l(\Gamma_l-\gamma_{ml})}
=\frac{\gamma_{ml}\Gamma_n}{\Gamma_l(\Gamma_m+\Gamma_n-\gamma_{mn})}.$$

At $G_3=0$ equations (1) and (2) convert in solutions for $\Lambda$ and $V$ schemes
$$r_{2}=i \frac{G_{2}}{P_2}\cdot \frac{\Delta r_2-g_3\Delta r_1}{1+g_2},\
r_{4}=i\frac{G_{4}}{P_4}\cdot \frac{\Delta r_4-g_1\Delta r_1}{1+g_4}.\eqno(4)$$
According to [22, 23, 27, 30], it is convenient to classify the effects of the strong
radiation resonant to an adjacent transition, as (i) population saturation (formulas
(3)), (ii) dynamic splitting of the resonance for a probing field (or splitting of a
common level, i.e., the ac Stark effect, denominators in formulas (4), and (iii)
nonlinear interference effects (NIEF) (the terms in the numerators of (4)). The two
last effects are conditioned by quantum coherence.
\section{ Difference in pure emission and absorption spectra  due  to \\ the
nonlinear  interference  effects:  inversionless  amplification,  \\ resonantly
amplified  refraction   in  the   absence  of  absorption, \\ and  laser-induced
transparency} A light emitted of absorbed, e.g., at the frequency $\omega_2$ whose
power is proportional to $Re( iG_2^*r_2)$, can be considered as a difference between
pure emission (a term proportional to $r_g$) and pure absorption (other terms in
formulas (1) and (4)). The two constituents are positive, but differently depend on
detuning due to the NIE. Thus, a sign alternation arises in spectral line contour,
resulting in the inversionless amplification. This was emphasized in [24, 27] (see
also [22, 23, 30, 36]). Optimum conditions for the inversionless amplification in a
uniformly broadened three-level system were analyzed in [28-30] in detail.  The
refraction index at frequency $\omega_2$ is defined as $Im(-iG_2^*r_2)$ and,
generally, the laser-induced minimum (including zero) absorption can coincide with
the resonant refraction index maximum [11, 40].  As is emphasized in [22-30, 36], the
splitting effect and the NIE as a whole, varying the spectral line shape and causing
the difference in pure emission (spontaneous or induced) and absorption spectra, does
not vary its integral intensity, which is defined only by saturation effects, $$ \int
{d\Omega_2 {\rm Re}(-ir_2/G_2)}=\Delta r_2, \quad \int {d\Omega_4 {\rm
Re}(-ir_4/G_4)}=\Delta r_4. \eqno(5) $$

Thus, namely NIE lead to the coherent population trapping, electromagnetically induced
transparency and the inversionless amplification, e.g., at the transition $gn$ (or
$ml$), when the second  terms in nominators of (4) become equal or begin to exceed
$\Delta r_2$ (or $\Delta r_4$). It is seen from the density matrix equation that the
considered effects are finally defined by the coherence at transitions $gm$ and $ln$
($r_{32}$ and $r_{12}$), induced jointly by probing and strong fields.
\subsection{Inversionless amplification of the probing wave}
Now we enlarge on the problem, what are the elementary processes usually defined by
the perturbation theory, which contribute into absorption (amplification) in the
analyzed cases. For instance, let us consider the absorption index $\alpha(\Omega_4)$
at frequency $\omega_4>\omega_1$ (Fig. 1) and $E_3=0$ normalized to its maximum
$\alpha^0(0)$, in the absence of all strong fields. From formulas (4) we find $$
\frac {\alpha(\Omega_4)}{\alpha^0(0)}= {\rm Re} \left \{ \frac {\Gamma_4[\Delta
r_4-g_1\Delta r_1]}{P_4\Delta n_4 (1+g_4)} \right \}. \eqno(6) $$ Further we consider
the two following cases. (i) Great yields from one-photon resonances $$ |\Omega_1|
\approx |\Omega_4| \gg \Gamma_1, \Gamma_4, \quad |g_4| \ll 1, |g_1| \ll 1, \quad P_4
\approx i\Omega_4, P_1 \approx i\Omega_1 \approx i\Omega_4. $$ Formula (6) takes on
the form $$ \frac {\alpha(\Omega_4)}{\alpha^0(0)} \approx \frac {\Gamma_4^2\Delta
r_4} {\Omega_4^2\Delta n_4}- {\rm Re} \left \{ \frac {\Gamma_4(\Delta r_4 g_4+\Delta
r_1 g_1)}{i \Omega_4 \Delta n_4} \right \} \approx \frac {\Gamma_4^2\Delta
r_4}{\Omega_4^2\Delta n_4}- \frac {\Gamma_4
\Gamma_{14}}{\Gamma_{14}^2+(\Omega_4-\Omega_1)^2} \frac {|G_1|^2 (\Delta r_1-\Delta
r_4)}{\Omega_4^2 \Delta n_4} $$ $$ = \frac {\Gamma_{lm}^2(r_l-r_m)}{(n_l-n_m)
\Omega_4^2} - \frac {\Gamma_{gm} \Gamma_{lm}}{\Gamma_{gm}^2+(\Omega_4-\Omega_1)^2}
\frac {|G_1|^2 (r_m-r_g )}{\Omega_4^2(n_l-n_m)}. \eqno(7) $$ The two last co-factors
in (7) describe the Raman scattering and arise from the nominator (NIE) and
denominator in (6). It ensues from (7) that the population inversion of initial and
final unperturbed states ($r_m=n_m>r_g$) is required for the probing field
amplification.

(ii)
Resonance $\Omega_1=\Omega_4=0$.

The amplification and transparency conditions have the form $$ g_1 \Delta r_1 \geq
\Delta r_4, \quad \frac {|G_1|^2}{\Gamma_{lg}\Gamma_{gm}} (r_l-r_g) \geq r_l-r_m.
\eqno(8) $$ As it follows from (8), the amplification, due to NIE, does not require
population inversion between initial and final states. The lower is the relaxation
rate at a two-photon transition as compared to the coherence relaxation at coupled
one-photon transitions, the more favorable are conditions for the inversionless
amplification. An optimum strong field intensity is defined by the common level
splitting into two quasilevels, which reduces the interference and, hence, the
amplification at the $ml$ transition center. The population difference saturation at
the strong field transition also reduces the system coherence. There is an optimum
relationship between the initial population differences $\Delta n_4$ and $\Delta
n_3$, created by an additional noncoherent excitation. Optimum conditions for the
inversionless amplification and transparency for opened and closed systems are
analyzed in [28-30] in more detail.

\subsection{Three-level \ system \ in  \ strong \ fields: \
inversionless \ amplification \ for \ the \  strong \ wave}

Above expressions can be easily generalized to the case of inversionless
amplification of the strong fields which can drive a quantum system. This case is of
interest in connection with creation of "laser without population inversion". For
certainty, let us consider the interaction of two strong fields $E_3$ and $E_4$ (Fig.
1). Taking the strong field $E_4$ effects in density matrix equations into account,
the set of equations can be reduced to an algebraic. The solution has the form $$
r_4=i \frac {G_4}{P_4} \frac {(1+u_2^*)\Delta r_4-v_1 \Delta r_3}{1+v_4+u_2^*}, \quad
r_3=i \frac {G_3}{P_3} \frac {(1+v_4^*)\Delta r_2-u_3 \Delta r_4}{1+v_4^*+u_2},
\eqno(9) $$ where $$ u_2= {|G_4|^2}/{P_3 P_{43}^*}, \quad u_3= {|G_4|^2}/{P_4^*
P_{43}^*}, $$ other notations are the same.

It is seen comparing (4) and (9) that, apart from the population difference
saturation, a growth in the amplified wave intensity makes more difficult to achieve
the conditions for inversionless amplification and self-transparency at the line
center (factors $(1+u_2^*)$ and $(1+v_4^*)$ in the nominators), as well as reduces
the gain due to an additional resonance splitting (the additional term in
denominators). An extended analysis of this problem, accounting for the saturated
populations, will be published elsewhere.
\subsection{ Inversionless  amplification  and  resonantly  amplified
refraction  in  the  absence  of  absorption  in  sodium  vapor: a simple experiment}
Currently a small number of experiments contrasts to a flux of theoretical
publications. The most experiments concern with coherent excitation of a doublet or a
set of neighboring sublevels in the short-pulse mode, as well as with accompanying
interference effects. In [41-43] there was proposed a design, comprehensive
theoretical grounds, and estimations for a possible experiment on simultaneous
observation of inversionless amplification and absorptionless resonant refraction.
This was a scheme of interfering two-quantum transitions induced by an auxiliary
field in the uniformly broadened three-level system with collisions. Such an
experiment is of interest due to minimized accompanying processes. Meanwhile, this
simplest model enters more complex experimental schemes.

Let us consider again the energy level diagram displayed in Fig. 1. We assume the
level $n$ to be a ground one and the field $E_1$ and $E_4$ to be absent. Thus, we
separate the $V$-shaped three-level configuration $g-n-m$. The strong field $E_3$ at
frequency $\Omega_3$ couples the levels $m$ and $n$. The weak field at frequency
$\omega_2$ probes the transition $gn$. Using (3) and (4) we derive the absorption
$\alpha_2$ and refraction $n_2$ indices at frequency $\omega_2$, (see also [28-30]),
$$ \alpha_2(\Omega_2)=\alpha_2^0(0) {\rm Im} f(\Omega_2, |E_3|^2), \eqno(10) $$ $$
\Delta n(\Omega_2)= n(\Omega_2)-n(\Omega_2)^{(nr)}= \delta n_2^0 {\rm Re} f
(\Omega_2,|E_3|^2, \eqno(11) $$ $$ f(\Omega_2,|E|^2)= -i \frac {\Gamma_{gn}}{\Delta
n_{ng}} \frac
{\Gamma_{gm}+i(\Omega_2-\Omega_3)(r_n-r_g)-ir_{mn}G_3}{[\Gamma_{gn}+i\Omega_2][\Gamma_{gm}+i(\Omega_2-\Omega_3)]+|G_3|^2},
\eqno(12) $$ where $\alpha_2^0(0)$ is the absorption (or amplification, depending on
the population difference $\Delta n_2$ sign) at the spectral line center if the
strong field $E_3$ is turned off, $\delta n_2^0$ is the maximum contribution of
transition $ng$ to the refraction index at $E_3 = 0$, and $n(\Omega_2)^{nr}$ is the
linear contribution of all other nonresonant levels.

As was noted above, the NIE leads to inversionless amplification and is created there
by the coherence at the transition $gm$. The coherence is induced by the strong field
(factor $r_{mn}$) in combination with the probing field. The greater is
$|G_3r_{nm}/\Gamma_{gm}|$ as compared to $r_n-r_g$, the more pronounced is the effect,
$$ r_{mn}=- \frac {i G_3(r_n-r_m)}{\Gamma-i\Omega_3}, \eqno(13) $$ hereafter
$\Gamma\equiv\Gamma_3$.  At $\Omega_3 = 0$ the absorption (amplification) maximum
corresponds to $\Omega_2 = 0$, hence, $$ f(0)= \frac{r_n-r_g-(r_n-r_m)|G_3|^2/ \Gamma
\Gamma_{gm}}{(1+|G_3|^2/\Gamma_{gm} \Gamma_{gn})\Delta n_2}. \eqno(14) $$ Thus, even
at $(r_n-r_g)>0$ and $(r_n-r_m)>0$, a negative absorption, i.e. amplification, could
take place, if $$ {|r_n-r_m||G_3|^2}/{\Gamma \Gamma_{gm}}> |r_n-r_g|. \eqno(15) $$ The
lower is the coherence relaxation rate $\Gamma_{gm}$ at two-photon transition $gm$ as
compared to the coherence relaxation at coupled one-photon transitions, the more
favorable are conditions for inversionless amplification. At
$|G_3|^2\gg\Gamma_{gm}\Gamma_{gn}$ a splitting of the level $n$  into two
quasi-levels significantly reduces interference and, hence, amplification at the
transition $gn$ center. There is also an optimum relationship between saturated
population differences at the interacting transitions. It depends on the strong field
intensity and the relation between initial population differences $\Delta
n_2=n_n-n_g$ and $\Delta n_3=n_n-n_m$ [28-30], created by an additional noncoherent
radiation.  To vary this relation in a wide range, we proposed in [41-43] to use
alkali atoms placed into a high-pressure buffer gas.  The strong field couples
$P_{3/2}$ and the ground $S$ - level. A fast collisional exchange furnishes
population transfer from $P_{3/2}$ to the lower level $P_{1/2}$. For simplicity, it
can be believed that the population distribution over the fine structure levels is
Boltzmann's one due to collisions. Thus, it becomes possible to vary the population
difference at the probing transition in a wide range varying the strong field
intensity and buffer gas pressure. Due to the saturation of $P_{3/2} - S$ transition,
even the population inversion at the $P_{1/2} - S$ becomes possible (similar to a
ruby laser). Hence, $P_{1/2} - S$ can be chosen as a probing transition. The
population inversion was experimentally observed by a similar scheme in the mixture
of sodium and helium vapors [44].

Collisions play a double part, i.e., on the one hand they considerably
worsen coherence, on the other hand the population transfer due to
collisions furnishes simple control and optimization of the population
differences at the coupled transitions. Moreover, a wide collisional
broadening allows one to neglect the nonuniformity of interaction with
atoms due to the Doppler effect, hyperfine splitting, and some other
processes. This makes an experiment be governed by the simplest theoretical
model. By estimations and numerical examples for sodium atoms, now we show
that inversionless amplification and absorptionless resonant refraction can
be significant under proposed experimental conditions, in spite of
collisions reducing the coherence.

We will discuss the concrete $D-1$ and $D_2$ transitions in sodium. According
to [44], we write the kinetic equations of level populations
$$
(\Gamma_m+\nu_{mg})r_m-\nu_{gm}r_g-P=0,
$$
$$
P= \frac {2|G_3|^2\Gamma(r_n-r_m)}{\Gamma^2+i\Omega_3^2}, \eqno(16)
$$
$$
(\Gamma_g+\nu_{gm})r_g-\nu_{mg}r_m=0,
\quad
\Gamma_m r_m+\Gamma_g r_g-P=0, \eqno(17)
$$
$$
r_m+r_n+r_g=N, \eqno(18)
$$
where $\nu_{gm}$ and $\nu_{mg}$, are the frequencies of collisions transferring
populations, $\Gamma_g^{-1}$ and $\Gamma_m^{-1}$ are the lifetimes of
relevant levels. From (16)-(18) we find
$$
r_n-r_m= \frac {N}{1+\kappa},
\quad
r_n-r_g=\frac {N}{1+\kappa} \left [
{\kappa} \frac {\nu_{mg}-(\nu_{gm}+\Gamma)}{\nu_{mg}+2(\nu_{gm}+\Gamma)} -1
\right ], \eqno(19)
$$
where
$$
{\kappa} = \frac {2|G_3|^2 \Gamma}{\Gamma^2+\Omega_3^2} \frac
{\nu_{mg}+2(\nu_{gm}+\Gamma)}{\Gamma_g\nu_{mg}+\Gamma_m(\nu_{gm}+\Gamma)}.
\eqno(20) $$
Let us assume that $\Gamma_g\approx\Gamma_m$ and a buffer gas pressure is high that
$(\nu_{mg}-\nu_{gm})\gg\Gamma_{gm}$. Taking   under   given   conditions
$\nu_{gm}=\nu_{mg}\cdot\exp(-\Delta E/k_bT)$, where $\Delta E=E_m-E_g$, is
the fine splitting energy, $k_b$ is the Boltzmann constant, and T is the
temperature, we get
$$
r_n-r_g=\frac {N}{1+\kappa} \left [ {\kappa} \frac {1-\exp (-\Delta E/k_B T)}{1+2\exp (-\Delta E/k_B T)}-1
\right ], \eqno(21)
$$
$$
{\kappa}= \frac {2|G_3|^2\Gamma}{\Gamma_m(\Gamma^2+\Omega_3^2)}
\frac {1+2\exp(-\Delta E/k_B T)}{1+\exp(-\Delta E/k_B T)}. \eqno(22)
$$
For sodium $\Delta E = 17.2$ cm$-^1$ and, at T = 550 K, estimations yield
$\Delta E/k_bT=4.3\cdot 10^{-2}$, $\kappa\approx 3|G_3|^2/\Gamma\Gamma_m
\approx 9\lambda^3I/64\pi^3\epsilon_0\hbar c\Gamma$,
$$
r_n-r_g=\frac {N}{1+\kappa} \left [  1.310^{-2}\kappa-1 \right ]. \eqno(23)
$$
where $\lambda$ and $I$ are the strong field wavelength and energy flux
density, $\epsilon_0$ is the dielectric constant of vacuum.  From (13),
(14), (20), and (21), it is seen that principally attainable inversionless
amplification rises as $\Delta E$ grows (e.g., in $K$ and $Rb$).

The inelastic collision cross section in sodium and helium for the
transition $3P_{3/2} - 3P_{1/2}$ is $\sigma_{mg}\approx 4\cdot10^{-15}$
 cm$^2$.  At T=550 K and helium atmospheric pressure, estimations yields
$\nu_{mg}=N_{He}\bar v\sigma_{mg} \approx7.5\cdot 10^9$ s$^{-1}$. Since
$\Gamma_g\approx\Gamma_m\approx 6.2\cdot10^{-7}$ s$^{-1}$, the validity conditions
for approximation (21) are satisfied. Using data [45] for the collisional broadening
D of sodium and helium lines, we estimate the collisional halfwidth as $\Gamma\approx
5\cdot10^{10}$ s$^{-1}$, which exceeds the Doppler's width of this transition,
$\Delta\omega_D/2=4.7\cdot10^9$ s$^{-1}$ ($\Delta\nu_D/2 = 0.75$ GHz).  For our
conditions, we have $$ \frac {|G_3|^2}{\Gamma\Gamma_{gm}}\approx \frac
{|G_3|^2}{\Gamma_{gn}\Gamma_{gm}}   \approx  \frac {\kappa\Gamma_m}{3\Gamma_{gm}}, $$
$\kappa\approx 5\cdot10^9 I$, where $I$ is expressed in $W\cdot \hbox{cm}^{-2}$.

For the radiation power of 0.1 W focused into the footprint $A=10^{-5}$ cm$^2$ (the
confocal parameter is $b\approx 1$ cm), we find $|G_3|\approx 3.6$ GHz,
$\kappa\approx5\cdot10^2$, and $|G_3|^2/\Gamma\Gamma_{gm}\approx0.1$. These values
are optimal to vary the population difference around zero at the probing transition
$r_n-r_g$. Above estimations for the intensity 1-10 kW$\cdot$cm$^2$ required to change
noticeably the line shape agree well to the experiment with a change in the
population difference ratio at the coupled transitions (44]. Inversionless
amplification at $r_n-r_g=0$ is estimated as
$\alpha_2(0)/\alpha_2^0(0)\approx-\Gamma_m/3\Gamma_{gm}$. Accepting
$\Gamma_{gm}\approx\nu_{mg}$, we find a value about 0.3\% of the absorption in the
absence of strong field. It is seen that this value is very sensitive to the
coherence decay rate at the transition $gm$.

Absorption-amplification spectral line shapes, frequency interval positions and
halfwidths can be controlled, as analyzed in [29,30]. The line shape is very sensitive
to the strong field intensity and frequency detunings from resonance. The
amplification halfwidth increases and maximum decrease as the strong field intensity
grows above a certain value. The population difference saturation at the strong field
transition and the common energy level splitting reduce the inversionlesa
amplification, so that it should be optimized by an appropriate choice of the strong
field intensity and detuning. In our case the inelastic collision frequency is the
important optimization parameter. The refraction index (dispersion) is described by
$\hbox{Re}\,f(\Omega_2,|E_3|^2)$. In the framework of proposed experiment the
absorption-amplification coefficient and refraction index shapes can be controlled so
that the refraction maximum falls within the spectral interval of vanishing
absorption.

Thus, the considered model of three-level system with the interference controlled by
collisions makes it possible the amplification without population inversion and the
resonantly amplified refraction at vanishing absorption. Exact formulas are presented
to analyze optimum experimental conditions. Collisions destroying the coherence
reduce the effects, as compared to atomic beams. However, this decrease is comparable
to the effect of Doppler broadening in metal vapors. Advantages of the proposed
experimental design are a simplicity and the possibility to control populations at
the coupled transitions and to avoid interfering effects. This makes experiment
adequate to a simple theoretical model. An experiment on inversionless amplification
in the continuous mode in potassium vapor using collisional population of the upper
level of probing transition was carried oat in [46], however the four-level
configuration contribution was significant there. The latter was conditioned by the
transitions between hyperfine splitting sublevels and the incomplete overlapping of
optical transitions due to insufficient collisional broadening.

\section{ Coherence \ and \ frequency \ mixing:
multiple \ resonances \ at \ the \ condition \ of \ induced \ transparency}

A nonlinear-optical response sharply increases as interacting wave frequencies
approach one - and multiphoton resonances. This reduces required intensities of
initial fields down to the values corresponding to cw lasers [47-51]. However, due to
resonant absorption of primary and generated waves, there arise limitations from
above onto the atomic density. Quantum coherence alternatively manifests itself in
various optical processes.  In particular, as was shown by an example of bound-free
transitions [31, 32, 36], the absorption decrease can be not accompanied by an
effective nonlinear susceptibility decrease on frequency mixing and varies the
refraction index in another way.

Recently, an interest is growing to control matter - optical properties via the
quantum coherence effects, especially promising for shortwave generation [7, 52, 53].
Therewith, an accent is on the wave conversion at frequency mixing under the
condition of resonance with an absorbing transition between discrete levels and only
by the generated field. In [54] a scheme of totally resonant multiphoton interaction
was proposed, in which the quantum transition coherence and interference suppress
absorption of both primary and generated field. Therewith the atomic nonlinear
susceptibility is not subject to a significant destructive interference and rises by
many orders of magnitude due to simultaneous multiphoton and one-photon resonances.
Now available great atomic concentration rises additionally the nonlinear-optical
response of medium and yields new spectral dependencies conditioned by local field
effects (to be considered further).
\begin{figure}
  \centering
\includegraphics[width=.3\textwidth]{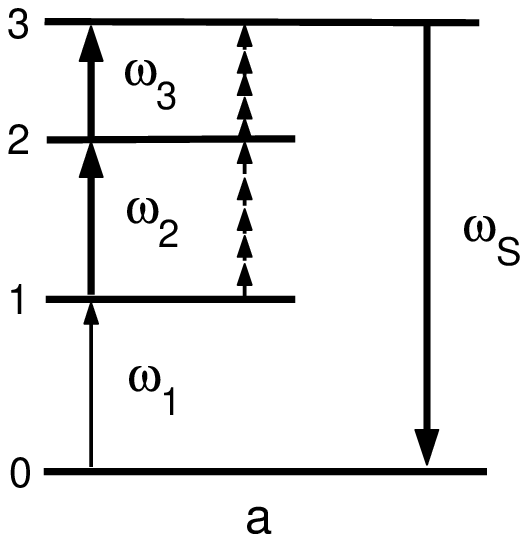}
\includegraphics[width=.3\textwidth]{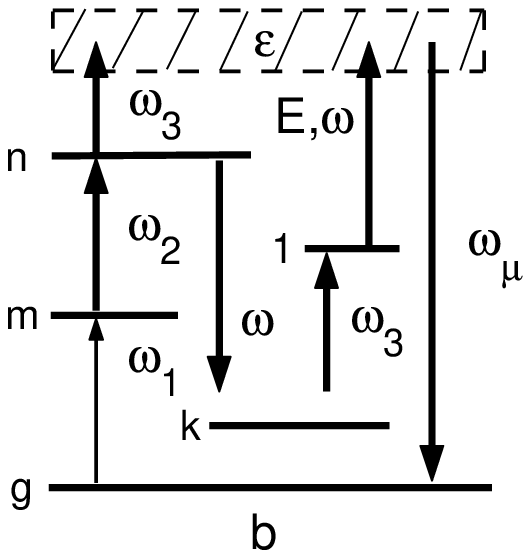}
\includegraphics[width=.3\textwidth]{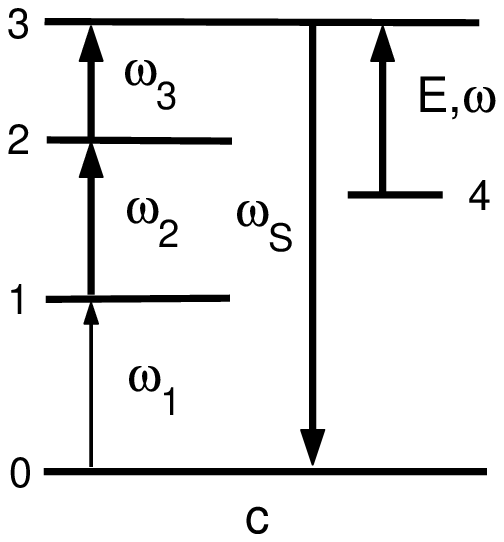}\\
  \caption{Frequency mixing enhanced by induced transparency: (a) triple
resonance (levels 2 and 1 as well as 3 and 2 are coupled by strong one- or multiphoton
interactions, levels 0 and 1 are coupled by a weak field), (b) autoionization-type
resonances induced in spectral continuum by strong fields $E$ and $E_3$, and (c)
three-photon resonant four-wave mixing increased by the additional strong field $E$.}
\end{figure}

In this section we consider several qualitative effects conditioned by
intratomic coherence and possible totally resonant four-wave interaction
under low absorption of both generated and initial fields. We show also the
possibility of significant efficiency of generation. The results are easily
extended to nonlinear-optical processes of higher order. Now we turn to the
energy level diagram shown in Fig.2a [54].

The strong fields $E_3$ and $E_2$ at frequencies $\omega_3$ and $\omega_2$ couple
nonpopuiated levels 3, 2 and 2, 1, respectively. The fields $E_1$ and $E_s$,
generated at frequencies $\omega_1\approx\omega_{10}$ and
$\omega_s=\omega_1+\omega_2+\omega_3$ are assumed to be weak and not changing the
level populations. The latter fields are considered only in the lowest order of
perturbation theory. The absorption coefficient and refraction index at frequencies
$\omega_1$ and $\omega_s$, as well as the nonlinear polarization generating the wave
at frequency $\omega_s$ are defined by real and imaginary parts of effective linear $$
\chi_1(-\omega_1;\omega_1)=(\chi_1^0/P_{01})f_1, \quad
\chi_s(-\omega_s;\omega_s)=(\chi_s^0/P_{03})f_s,   \eqno(24) $$ and nonlinear $$
\chi^{NL}(-\omega_s;\omega_1+\omega_2+\omega_3)=(\chi_0^{NL}/P_{01}P_{02}P_{03})f
\eqno(25) $$ susceptibilities, which, in turn, are proportional to the pre-exponentiat
factors $r_i$, and $\tilde r_i$ of the corresponding components of nondiagonal
density matrix elements (see the similar equations of Sec. 2). Here $\chi_1^0$,
$\chi_s^0$ and $\chi_s^{NL}$ -- are the resonant susceptibilities at negligibl $G_2$
and $G_3$. The factors $f_1$, $f_2$,
 and $f$ describe the strong field effects. Simple calculations by the density
matrix procedure similar to [30,36] yield $$ f_1=\left \{
1+g_2/P_{01}P_{02}[1+(g_3/P_{02}D_{03})]\right \}^{-1}, \eqno(26) $$ $$ f_s=\left \{
1+g_3/P_{03}D_{02}[1+(g_2/D_{02}D_{01})]\right \}^{-1}, \eqno(27) $$ $$
f=f_1[1+g_3/D_{03}P_{02}]^{-1}= \left [ 1+(g_2/D_{02}D_{01})+(g_3/D_{03}P_{02})
\right ] ^{-1}, \eqno(28) $$ where $$ P_{01}=1+ix_1, \quad P_{02}=1+ix_{02}, \quad
P_{03}=1+ix_{s}, $$ $$ D_{01}=1+iy_1, \quad D_{02}=1+iy_{02}, \quad D_{03}=1+iy_{s},
$$ $$ x_1=\frac {\omega_1-\omega_{10}}{\Gamma_{10}}=0, \quad x_{02}=\frac
{\omega_1+\omega_2-\omega_{20}}{\Gamma_{20}}=0, \quad x_s=\frac
{\omega_s-\omega_{30}}{\Gamma_{30}}=0, $$ $$ y_1=\frac
{\omega_s-\omega_3-\omega_2-\omega_{10}}{\Gamma_{10}}=0, \quad y_{02}=\frac
{\omega_s-\omega_3-\omega_{20}}{\Gamma_{20}}=0, \quad y_s=\frac
{\omega_1+\omega_2+\omega_3-\omega_{30}}{\Gamma_{30}}=0, $$ $$ g_2=
{G_2^2}/{\Gamma_{10}\Gamma_{20}}, \quad g_3= {G_3^2}/{\Gamma_{30}\Gamma_{20}}, $$ and
$\Gamma_{ij}$ are the uniform halfwidths of corresponding transitions. To analyze the
cases. when $E_s$, is not an independent probing field, we should put
$\omega_s=\omega_1+\omega_2+\omega_3$ and $D_{0i}=P_{0i}$.

Differences of the factors $f_1$, $f_2$, and $f$ from unity and between each other,
as well as their frequency dependency, are conditioned by two different coherence
initiation channels $\rho_{02}$ (two combinations of strong and weak fields, $E_1$,
$E_2$ and $E_s$, $E_3$) and their evolution as the field intensities rise. The
absorption coefficients are defined by imaginary parts of the relevant
susceptibilities, relative to which the considered processes manifest themselves as
resonance splitting and absorption minima. The generated wave power $P\propto
g_2g_3|\chi^{NL}|^2$ is defined by not only an imaginary part, but also by the real
part of $\chi^{NL}$.
 (A relative phase of the generated wave depends on their ratio.) Therefore,
the quantum coherence effects can be used to match the most important
generation conditions, i.e., significant decrease of absorption for all the
interacting fields without a noticeable decrease in atomic
nonlinear-optical response. Furthermore, these effects can be used to
increase additionally the generation efficiency improving the wave phase
velocity matching. The laser-induced spectral structures in the
susceptibility real parts (additional dispersion conditioned by the
coherence) enable such a matching by a slight detuning from the resonance
for $\omega_1$ or $\omega_s$.

An approach to the two resonances increases $|\chi^{NL}|^2$ by the factor $x_i^{-2}$,
equal to $10^6$ and more. The triple resonance gives a gain of the order of
$10^{18}$.  Due to the induced transparency, the atomic density N and, hence, the
power $P\propto N^2$ can be additionally increased by a few orders of magnitude.

These results are easily generalized to the case of higher order mixing.
For instance, if the 3-2 and (or) 2-1 transitions are multiphoton, the
generalization is achieved by substituting the one-photon Rabi frequencies
$G_2$ and $G_3$ with the corresponding matrix elements for multiphoton
transitions.

In a particular case, when the transparency is induced only at the frequency
$\omega_s$, by the wave $E_4$ not directly participating in conversion (Fig.  2c),
the relevant expressions are also deduced by a simple subscript substitution. It is
necessary to put $g_2=0$ and change subscripts in the formulas for $g_3$ and the
corresponding resonance denominator. This case is similar to considered in [31, 32,
36] for the transition to continuum (Fig.  26) (see also [52, 53]).

Thus, the intratomic coherence effects in strong electromagnetic fields
enable one to control absorption, refraction, and nonlinear-optical
generation spectral properties. In particular, the above choice of
transition diagram and interacting wave intensities make it possible to
gain a medium nonlinear-optical response increased by many orders,
meanwhile eliminating the initial and generated field absorption.

\section{ Atomic coherence: effects of a local field and the spectral line
nonuniform broadening}

\subsection{ Local field}
The local field acting on an atom substantially differs from the external field both
in value and phase, as the atomic density grows. This varies shapes of the spectral
lines conditioned by the quantum transitions' interference [55]. The effect is
revealed at particle concentration of the order of $10^{17}$ cm$^3$. Let us show by a
simple example that qualitatively new spectral dependencies can arise in above
problems due to the local field formation and drastic variation in the field of
additional strong laser radiation at an adjacent transition.

According to a conventional (but approximate) concept (see, e.g. [56, 57]), the local
$E_L$ and external $E$ fields in isotropic media are related by the simple formula $$
E_L=E+\frac {P}{3\epsilon_0}  . \eqno(29) $$ The medium polarization $P$ to a linear
approximation can be presented by $P = \epsilon_0N\alpha E_L$, where $N$ is the
atomic concentration,
 $\alpha$ is the microscopic (atomic) polarizability, and $\epsilon_0$ is
the dielectric constant of vacuum.

One of the consequences from (29) is the Clausius-Mossotti equation [56, 57] relating
polarizability $\alpha$ to the dielectric constant $\epsilon_0$ of material $$
\epsilon=1+LN\alpha, \eqno(30) $$ where the local field factor
$L=(\epsilon+2)/3=(1-\alpha N/3)^{-1}$ shows how the local field differs from the
external one. The former plays an important part in linear and nonlinear optical
phenomena (see, e.g. [56-59]. In spite of an approximate nature of formula (29), the
authors of [58] showed that it well describes linear and nonlinear responses of the
dense atomic gas.

Let us consider the interaction of two optical fields with three-level
systems of the cascade or $\Lambda$-schemes. Therewith, one field is strong
and (for simplicity) interacts with the transition between nonpopulated or
equipopulated levels.
According to the classification of resonant nonlinear processes (see Sec.
2), only the resonance splitting effect would be observed for the probing
field in this case. For certainty we consider the $n-m-l$ $\Lambda$-scheme
(Fig. 1), where the levels $m$ and $n$ are nonpopulated, but the state $l$
 is
ground. The strong field of frequency $\omega_3$ and amplitude $E_3$ acts
at the $nm$ transition, the weak (probing), wave of frequency $\omega_4$
 and amplitude $E_4$ acts
at the $lm$ transition. The set of equations for this problem differs used in
Sec. 3 only by the substitution of $E_4$ by $E_{4L}$. Such an approach is
widely used in the local field theory [56-59].

For the macroscopic complex polarization $P_4(\omega_4)=d_{ml}\rho_{lm}N$ using
formulas of Sec. 3 we find $$ P_4(\omega_4)=\epsilon_0\chi_4(\omega_4)E_4, \quad
\chi_4(\omega_4)= \chi_4^0 f(\omega_4), \eqno(31) $$ $$ \chi_4^0=i \frac
{N|d_{ml}|^2}{\epsilon_0\Gamma_{lm}}, \quad f(\omega_4)= \frac {\Gamma_{lm}P_{43}}
{(P_4-i\delta_{4L})P_{43}+|G_4|^2}, \eqno(32) $$ where $\chi_4$ is the macroscopic
susceptibility at the probing field frequency $\omega_4$ in the presence of strong
one at the frequency $\omega_3$, $\chi_4^0$ is the susceptibility in the absence of
the latter, and $f(\omega_4)$ is the form-factor. The parameter $$
\delta_{4L}=|d_{ml}|^2 \frac {N}{3\epsilon_0 \hbar} \eqno(33) $$ appears as the
transition $lm$ frequency shift conditioned by the concentration rise (local field).
It is substantial that in so doing the two-photon and strong field transition
frequencies are not varied. As a result, the local field effect is not reduced to
redefining the detuning of the weak field resonance, but qualitatively changes the
whole spectral line shape, if this factor becomes comparable to the resonance width.
At $d_{ml}$=1 Db and $N=10^{23}$ m$^{-3}$ we estimate this shift as
$\delta_{4L}=8\cdot10^{11}$ s$^{-1}$, which can be comparable to characteristic shock
widths of resonances. For instance, when the shock width is defined by resonance
exchange (self-broadening) and significantly exceeds a natural one, $\Gamma_{ml}$ has
the form (see, e.g. (58,59] and references therein) $$ \Gamma_{ml}\approx |d_{ml}|^2
\frac {N}{6\epsilon_0 \hbar}, $$ whence it follows that the ratio
$C_4=\delta_{4L}/\Gamma_{ml}$ can approach two in this case.

Peculiarities of the local field, exhibited in spectral nonlinear
interference dependencies of absorption, are analyzed in [55]. The red
shift, entering all the resonant denominators, and hence, not leading to a
simple redefinition of the detuning from one-photon resonance, changes
qualitatively the spectral dependencies as the absorbing particle
concentration rises. Thus, the local field strongly varies spectral
dependencies of the probing field absorption in the presence of a strong
field at the adjacent transition.

The factor $L_4=E_{4L}/E_4$ characterizing the difference between local and
external fields by value and phase can be presented in the form
$$
L_4=1+iG_4f(\omega_4). \eqno(34)
$$
Whence it follows that this difference also increases as the parameter $C^4$
rises. Varying the strong field intensity and frequency, the local field
spectral dependence can be also significantly varied.

The resonant exchange also shifts the frequency of the groond-to-excited
state transition. This shift is proportional to atomic concentration, but
is usually two- or threefold shorter than the broadening and can be often
neglected.

The acquired results are genual and applicable to other interaction
schemes. In the case of cascade transitions under the conditions of zero or
equal level populations at the strong field transition, the results are
found by a simple redefinition of corresponding quantities.

Since the local field acting on an atom can substantially differ by value
and phase from the external field as atomic density rises, this drastically
changes spectral dependencies not only for absorption and refraction, but
also for generating nonlinear poiarizations [54]. Similar, it can be shown
using the Lorentz-Lorentz approximation that the local field induces red
shifts in the resonant denominators at the allowed transitions for
effective nonlinear susceptibilities (Sec. 4). Thus, a supplementary
opportunity appears to increase efficiency sharply by controlling
concentration, initial field intensities, and detuninga from resonances at
the conditions of multiple resonances, induced transparency, and phase
matching. The considered local field effects should be taken into account
when designing and interpreting experiments.

\subsection{ Constructive and destructive interference as a consequence of
atomic velocity distribution: nonuniform broadening of spectral lines}

As it was already mentioned, the laser-induced coherence contribution into spectra
can be constructive or destructive depending on a detuning sign. Therefore,
inversionless amplification and resonance splitting in gases can significantly differ
from those for stationary atoms at the Doppfer broadening prevailing above uniform
one. Nevertheless, as was shown in [22, 23, 27, 30, 60, 61], spectral profiles with
alternating signs are induced sometimes even in this case. In the relatively weak
laser fields the NIE manifest itself in narrow spectral structure variations within
the wide Doppler's contour.  This structure is anisotropic, i.e., depends on the
angle between the strong and probing field wave vectors, as well as on collisions
changing atom velocities. Destructivity or constructivity effect of  the Maxwell
velocity distribution depends on the position of probing wave frequency compared to
the strong-field frequency. The probing field transition line is being deformed as a
whole as the strong field transition uniform broadening or the wave intensity rise.
Formulas of averaging over velocities for a number of cases are presented in [27, 30,
60, 61].

Since the dependence of responses on detuning signs often has a
sign-alternating character at coherent interaction, some interfering
components can vanish when averaged over velocities. We present an example
showing the effect of nonuniform broadening on coherent interaction. Let us
consider the four-wave mixing in the lowest order of perturbation theory
(Fig. 1). From the solutions to equations for $\bar r_4$ (Sec. 2) we deduce
the nonlinear susceptibility at frequency $\omega_4=\omega_1-\omega_2+\omega_3$
$$
\chi(\omega_4=\omega_1-\omega_2+\omega_3)=
\frac {iK}{\Gamma_{ml}+i(\Omega_1^{'}-\Omega_2^{'}+\Omega_3^{'})} \times \eqno(35)
$$
$$
\left \{
\frac {1}{\Gamma_{gm}+i(\Omega'_3-\Omega'_2)}
\left [
\frac {n_g-n_n}{\Gamma_{ng}-i\Omega'_2}+\frac {n_m-n_n}{\Gamma_{mn}+i\Omega'_3}
\right ]+
\frac {1}{\Gamma_{ln}+i(\Omega'_1-\Omega'_2)}
\left [
\frac {n_g-n_n}{\Gamma_{ng}-i\Omega'_2}+\frac {n_g-n_l}{\Gamma_{lg}+i\Omega'_1}
\right ]
\right \}
$$
where $K$ is the constant,
$\Omega'_1=\omega_1-\omega_{gl}-\bf{k}_1v=\Omega_1-\bf{k}_1v$ is the
detuning from resonance tailing the Doppler's shift into account, other
$\Omega'_i$ are the similar detunings depending on velocity, and $n_i$, are
the populations of corresponding levels, also depending on velocity.

As is seen from (35), all the terms, besides those proportional to $n_g-n_n$
as functions of velocity, have poles at the same complex semiplane.
Therefore, if the Doppler's shifts corresponding to the heat velocity $u$
much exceed uniform halfwidths of transitions, then only the polarization
components proportional to $n_g-n_n$ are nonzero after averaging over
velocities with the Maxwell distribution. The averaging result has the form
$$
\langle\chi\rangle_v=
\frac {iK\pi^{1/2} \exp
\left \{-(\Omega_2/k_2u)^2\right \}(N_g-N_n)}
{k_2u[\tilde \Gamma_1+i(\Omega_1-k_1\Omega_2/k_2)] [\tilde \Gamma_3+i(\Omega_3-k_3\Omega_2/k_2)]}, \eqno(36)
$$
where $N_g$, and $N_n$ are the integral over velocities unperturbed level
populations, $\Omega_4=\Omega_1-\Omega_2+\Omega_3$,
$\Omega_1=\omega_1-\omega_{gl}$, $\Omega_2=\omega_2-\omega_{gn}$,
$\Omega_3=\omega_3-\omega_{mn}$, $k_i=\omega_i/c$   and
$$
\tilde \Gamma_1=\Gamma_{nl}+(k_1/k_2-1)\Gamma_{ng}, \quad
\tilde \Gamma_3=\Gamma_{gm}+(k_3/k_2-1)\Gamma_{ng}. \eqno(37)
$$
Substituting $\Omega'_2$ by $-\Omega'_2$ in the sum
$\omega_4=\omega_1-\omega_2+\omega_3$ in the resonant cascade level
configuration, we find that all the poles appears at the same complex
plane. This significantly decreases the averaged susceptibility as compared
to the frequency subtraction scheme.

The effect of velocity distribution on the coherent four-wave mixing can be
described by the following way. For stationary atoms and a totally resonant
process, the nonlinear polarization is proportional to the factor
$\Gamma^{-3}$.  For a gas and the frequency subtraction at an available
intermediate level population, this factor is substituted by
$1/ku\Gamma^2$, i.e., decreased by $ku/\Gamma$. A more detailed analysis
[30] shows that for frequency summation or at $N_n - N_g = 0$, the
coherence and nonlinear polarization suppressed by the interference of
various velocity contributions yields this factor equal to $(ku)^{-3}$. In
other words, the susceptibility decreases by the factor of
$(ku/\Gamma)^{3}$ as compared to stationary atoms and by the factor of
$(ku/\Gamma)^{2}$  as to optimum conditions for frequency subtraction in
gases. Hence. the difference scheme was chosen for continuous four-wave
generation in the field of helium-neon laser in [47] as distinct from [48].

When only one resonance ($\omega_1-\omega_2\approx\omega_{ln}$) of the
Raman-type scattering presents, we have
$$
\chi^{(3)}\propto \frac {1}{\Omega_1\Omega_4}\exp \left [-\left ( \frac {\Omega_1-\Omega_2}{(k_1-k_2)u} \right )^2 \right ].
$$
In the absence, of resonancea (for both frequency summation and
subtraction), we find
$$
\chi^{(3)}\propto 1/\Omega_1\Omega_4(\Omega_1-\Omega_2).
$$
Since the NIE and mixing processes are defined by a common one source,
i.e., the coherence induced at a forbidden transition, dose effects occur
also at averaging over velocities.

Thus, the transition nonuniform broadening can significantly change the
 influence of various level population on the coherent processes, so that
 some level contributions prevail. Varying the level populations, strong
 field intensities and detunings, the amplification-absorption, refraction,
 and nonlinear polarization spectra can be controlled, thus considerably
increasing che generation yield.

\section{ Nonlinear \ interference \ effects \ at \ bound-free \ transitions:
laser-induced \ autoionization-type \ resonances \ in \ the \ continuum}

Nonlinear interference effects similar to those occurring at the discrete transitions
(including inversionless amplification and induced transparency) manifest themselves
also in continuous spectra, e.g., at the transition into an ioniza-tion continuum.
The corresponding theory was generalized in [30-32, 36, 62]. Similar phenomena for
crystal brands were considered in [29]. The laser-induced and autoionization-type
resonances theoretically analyzed in [31] were experimentally observed in [32], and
then the nonlinear interference processes became a subject of intense studies in the
context of laser-induced continuum structures (LICS), inversionless amplification and
electromagnetically induced transparency, first, at bound-free transitions (see, e.g.
[62, 64, 65]) and then also at discrete transitions (see [7-11, 40, 52, 53]).

Let us show the potentials to control simultaneously two LICS and to split discrete
resonances by strong electromagnetic fields to reduce absorption, correct phase
matching, and to improve nonlinear-optical short-wave generation techniques (Fig.
2b). The wave at frequency $\omega_1$ is weak, but those at frequencies $\omega_3$,
and $\omega$ are strong. We also take into account possible strong nonresonant
transitions to the discrete levels $k$.  The detunings $|\omega_1-\omega_{gm}|$,
$|\omega_1+\omega_2-\omega_{gn}|$ and $|\omega_1-\omega_3-\omega_{ln}|$ are assumed
to be much smaller than all other. Density matrix calculations similar to [31, 36]
yield the following expressions for nonlinear susceptibility
$\chi^{(3)}(\omega_{\mu}=\omega_1+\omega_2+\omega_3)$, the probing wave absorption
coefficients $\alpha(\omega_1)$ and $\alpha(\omega_{\mu})$ at the corresponding
frequencies, $$ \frac
{\chi^{(3)}(\omega_\mu=\omega_1+\omega_2+\omega_3)}{\chi^{(3)}_{0\mu}}= \frac
{K}{D_{gm}X}, \eqno(38) $$ $$ \frac {\alpha(\omega_1)}{\alpha_{01}}= {\rm Re} \left \{
\frac {1-g_{mn}/(D_{gm}X)}{D_{gm}}\right \}, \eqno(39) $$ $$ \frac
{\alpha(\omega_\mu)}{\alpha_{0\mu}}= 1-k_3\beta_l+\frac
{k_3\beta_l(y_l+q_{gl})^2}{1+y_l^2}- {\rm Re} \left \{ k_4g_{nn}A^2\frac
{(1-iq_{gn})^2}{Y} \right \}, \eqno(40) $$ where $\chi^{(3)}_{0\mu}$, $\alpha_{01})$
and $\alpha_{0\mu})$ are the relevant resonant values for negligible intensities of
all the fields, $$ K=1- \frac
{k_1\beta_l[(1-iq_{nl})(1-iq_{lg})]}{(1-iq_{ng})(1+ix_l)}, \eqno(41) $$ $$ A=1- \frac
{k_1\beta_l[(1-iq_{ln})(1-iq_{gl})]}{(1-iq_{gn})(1+iy_l)}, \eqno(42) $$ $$ X=
(1+g_{nn})\left [1+ix_n+ \frac {q_{mn}}{D_{gm}(1+q_{nn})}-k_2\beta_l\beta_n \frac
{(1-iq_{nl})^2}{1+ix_l} \right ], \eqno(43) $$ $$ Y= (1+g_{nn})\left [1+iy_n+ \frac
{q_{mn}}{p_{gm}(1+q_{nn})}-k_2\beta_l\beta_n \frac {(1-iq_{nl})^2}{1+iy_l} \right ],
\eqno(44) $$ $$ D_{gm}=1+ \frac {i(\omega_1-\omega_{gm})}{\Gamma_{gm}}, \quad
p_{gm}=1+\frac {i(\omega_\mu-\omega_3-\omega_2-\omega_{gm})}{\Gamma_{gm}}, \eqno(45)
$$ $$ x_l=\frac
{\omega_1+\omega_2+\omega_3-\omega-\omega_{gl}-\delta_{ll}}{\Gamma_{gl}+\gamma_{ll}},
\quad x_n=\frac {\omega_1+\omega_2-\omega_{gn}-\delta_{nn}}{\Gamma_{gn}+\gamma_{nn}},
\eqno(46) $$ $$ y_l=\frac
{\omega_\mu-\omega-\omega_{gl}-\delta_{ll}}{\Gamma_{gl}+\gamma_{ll}}, \quad y_n=\frac
{\omega_\mu-\omega_3-\omega_{gn}-\delta_{nn}}{\Gamma_{gn}+\gamma_{nn}}, \eqno(47) $$
$$ k_1=\frac {\gamma_{gl}\gamma_{ln}}{\gamma_{gn}\gamma_{nn}}, \quad k_2=\frac
{\gamma_{nl}\gamma_{ln}}{\gamma_{ll}\gamma_{nn}}, \quad k_3=\frac
{\gamma_{gl}\gamma_{lg}}{\gamma_{gg}\gamma_{ll}}, \quad k_4=\frac
{\gamma_{gn}\gamma_{ng}}{\gamma_{gg}\gamma_{nn}}, \eqno(48) $$ $$ g_{mn}=\frac
{|G_{mn}|^2}{\Gamma_{gm}\Gamma_{gn}}, \quad \beta_l= \frac {g_{ll}}{1+g_{ll}}, \quad
\beta_n= \frac {g_{nn}}{1+g_{nn}}, \eqno(49) $$ $$ g_{ii}=\frac
{\gamma_{ii}}{\Gamma_{gi}}, \quad g_{ij}=\frac {\delta_{ij}}{\gamma_{ij}}, \eqno(50)
$$ $$ \gamma_{ij}= \pi \hbar G_{i\epsilon}G_{\epsilon j} |_{\epsilon=\hbar\omega\mu}+
{\rm Re} \left \{ \sum_k \frac {G_{ik}G_{kj}}{p_{gk}}\right \}, \, \delta_{ij}=\hbar
P\int \frac {d\epsilon G_{i\epsilon} G_{\epsilon j}}{\hbar \omega_\mu-\epsilon}+ {\rm
Im} \left \{ \sum_k \frac {G_{ik}G_{kj}}{p_{gk}}\right \}, \eqno(51) $$ $P$ in (51)
designates the sign of the integral principal value. The factors $k_i$, take the
values $0\ge k_i\ge 1$ depending on a degree of degeneracy for continuum states
(unity for the nondegenerate states).

Formulas (38) and (40) generalize the expressions from [31, 36] to the case of
several strong fields.  Together with (39) these formulas show the possibility to
reduce absorption both for the primary and the generated fields. The absorption falls
exponentially as the medium length rises. Since the absorption coefficients, as
functions of frequency, and the ratios of squared module of nonlinear susceptibility
to these coefficients, do not coincide at certain conditions. These ratios define the
generation power in absorbing media. This power quadratically rises with growing
atomic concentration and the initial fields intensity under the condition of
optimized medium perturbation. Comparing (38) and (40) to corresponding formulas from
[31, 36] and Sec. 4, we see new interference spectral structures caused by the joint
action of strong fields $E_3$and $E$ (terms proportional to $\beta_n$ and $g_n$),
arising in nonlinear polarization and absorption (refraction). These nonlinear
resonances give additional opportunities for absorption nonlinear spectroscopy and
enhancement of efficiency for nonlinear-optical conversion to the short-wave spectral
range.
\section{ Nonlinear interference and relaxation}
Relaxation processes can extraordinarily exhibit the coherence effects. For instance,
the spontaneous relaxation coupling two transitions with close frequencies [64, 65]
can promote inversionless amplification even in the absence of strong fields. Let us
consider an example [47], when certain relaxation mechanisms and external dc fields
suppressing the destructive interference make  the resonant nonlinear-optical
interaction allowed. The experiment was carried out with the $He-Ne$ laser beam
($\lambda=1.52$ $\mu$m) resonant to the transition $2s_2-2p_4$ of Ne atoms. Upper and
lower levels contain three ($J_1 = 1)$ and one ($J_0= 0$) Zeeman sublevels,
respectively. The initial beam contained two linear ortogonaily polarized components
$E_1$ and $E_2$ with the frequency shift $\Delta=\omega_2-\omega_1$ much lower than
the natural transition width. The wave intensity at frequency $\omega_1$ was
significantly higher than at
 $\omega_2$.  The four-wave generation of $E_s$, arose at the frequency
$\omega_s=2\omega_1-\omega_2=\omega_2-2\Delta$ with the same polarization as in
$E_2$. The generation power sharply rose with growing collision frequency and
external $dc$ magnetic field. This effect can be explained in the following way.
\begin{figure}[h!]
\centering
\includegraphics[width=.28\textwidth]{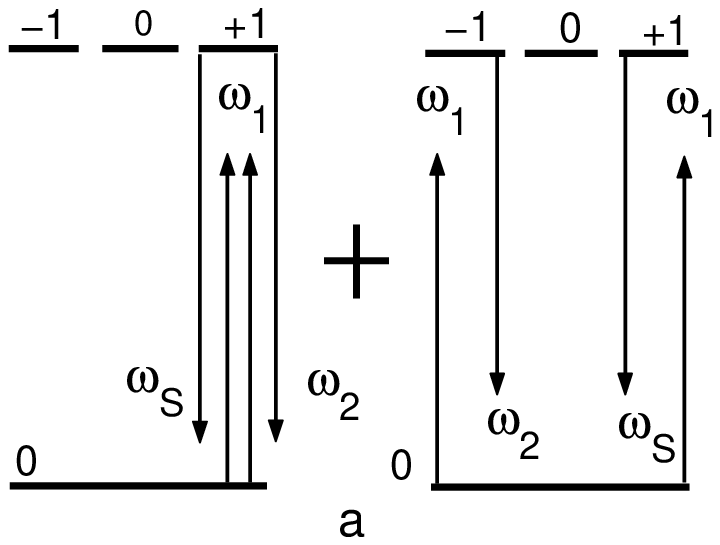}\qquad
\includegraphics[width=.16\textwidth]{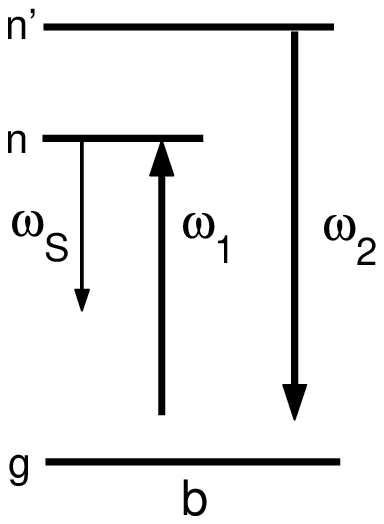}\qquad
\includegraphics[width=.16\textwidth]{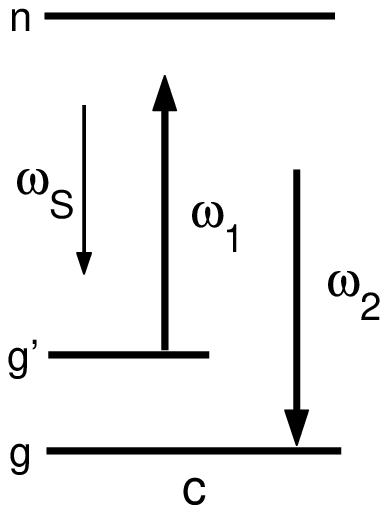}\\
\includegraphics[width=.15\textwidth]{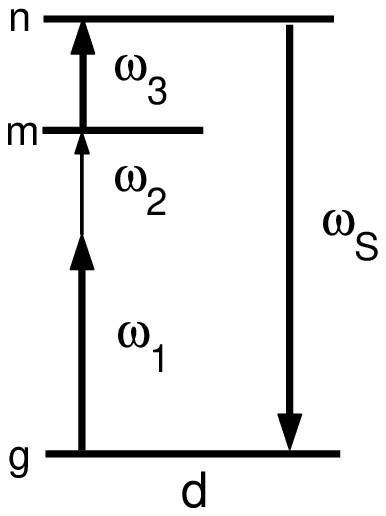}\quad
\includegraphics[width=.13\textwidth]{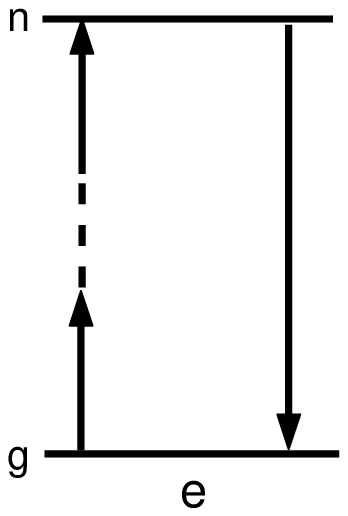}\quad
\includegraphics[width=.125\textwidth]{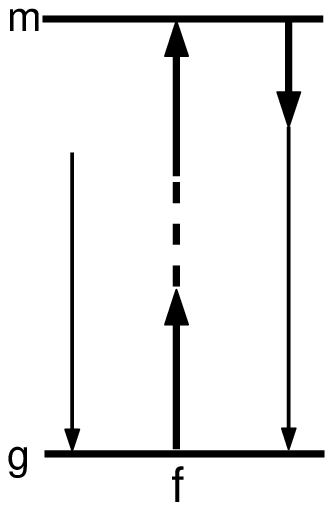}\quad
\includegraphics[width=.09\textwidth]{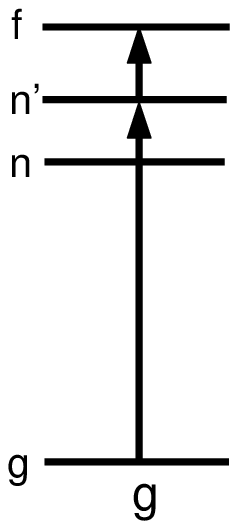}\quad
\includegraphics[width=.09\textwidth]{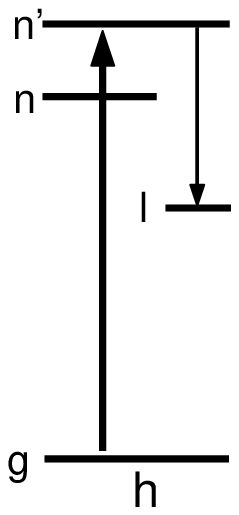}\\
  \caption{ Interfering oscillations in various resonant processes: (a)
destructive interference of four-wave mixing processes in the double $V$-shaped and
two-level configurations between the Zeeman sublevels of transition $2s_2-2p_1Ne$
[47], (b,c) interference of the doublet sublevels contributions into four-wave mixing
[66, 67], (d) up-conversion of the weak infrared ($\omega_2$) wave (fields $E_1$ and
$E_3$ are strong mainly destructive interference at the frequencies
$\omega_s-\omega_3=\omega_{ng}=\omega_1+\omega_2$ limits the conversion efficiency
[69], (e) interference of multiphoton and one-photon polarization components limits
the upper level population, (f) interference of nonresonant nonlinear polarization of
the seventh order and the resonant one of the ninth order are used to detect the
seventh harmonic generation at the resonant nonlinearity of the ninth order [70], and
(g,h) interference of the doublet component contributions into the two-photon and
nonresonant one-photon processes (spontaneous emission from the upper levels).}
\end{figure}
It is convenient to expand each wave and nonlinear polarization $P^{(NL)}(\omega_s)$
into two circular components $P_+^{(NL)}(\omega_s)$ and $P_-^{(NL)}(\omega_s)$ .
These components contain two terms. One describes the four-wave mixing with the same
polarization at two-level subsystems, while another does the radiation with contrary
polarizations at three-level Zeeman subsystems (Fig. 3a).

At such a choice of polarizations, these two contributions interfere destructively
and totally suppress one another, if relaxation rates for populations and quadripole
moments (alignment) are equal for the Zeeman sublevels in the upper electron state.
The spontaneous radiation trapping, anisotropic collisions, and/or the external
magnetic field violate the amplitude balance of destructively interfering components
of nonlinear polarization and induce the four-wave mixing. The magnetic field effect
on the two-level configuration is compensated for the Doppler shifts. The second
component of nonlinear polarization represents a double $V$-configuration (Fig. 3a),
which is removed from resonance by the magnetic field.

Now let us consider another example (Fig. 3a). The coherence induced at the
transition $n'n$, defining four-wave mixing $\omega_s=2\omega_1-\omega_2$, is given by
$$ \rho_{n'n}^{(2)}\alpha V_{n'g} \rho_{gn}^{(1)}+ \rho_{n'g}^{(1)} V_{gn}\alpha
\left [ \frac {1}{\Omega_2+i\Gamma_{n'g}}- \frac {1}{\Omega_1+i\Gamma_{ng}}\right ]
\frac {1}{\Omega+i\Gamma_{n'n}}= $$ $$ \frac
{1}{(\Omega_2+i\Gamma_{n'g})(\Omega_1-i\Gamma_{ng})} \left [ 1-i \frac
{\Gamma_{nn'}-\Gamma_{n'g}-\Gamma_{ng}} {\Omega+i\Gamma_{nn'}}\right ], \eqno(52) $$
where $\Omega_1=\omega_1-\omega_{ng}$, $\Omega_2=\omega_2-\omega_{n'g}$,
$\Omega=\omega_2-\omega_1-\omega_{n'n}$.  A spontaneous relaxation we have
$\Gamma_{ij}=(\Gamma_i+\Gamma_j)/2$ and the resonance $\Omega=0$ disappears.
Collisions violate the relevant frequency equality and induce this resonance,
suppressing the destructive interference [66, 67].

\section{ Conclusion}
The purpose of this paper is to show the diverse exhibition of nonlinear interference
effects in optics and to survey certain earlier works in this field. Interference
phenomena can play a governing part in numerous experiments on resonant nonlinear
optics [68]. Some such processes are sketched at Fig. 3 and explained in the figure
caption.

The author is cordially grateful to V.A. Ignatchenko, with whom he began to deal with
laser physics, to S.G. Rautian, who introduced him into nonlinear laser spectroscopy,
and to V.P. Chebotaev and S.A. Akhmanov, who taught to him many things  but are not
among living, unfortunately.

I greatly acknowledge also to my colleagues Im Thek-de, Yu.I. Heller, V.V. Slabko, and
V.G. Arkhipkin for their collaboration.

The author is also grateful to B. Wellegehausen, H. Welling, and G. zu Putlitz and
the German Research Society; L.J.P. Hermans and J.P. Woerdman and the Netherlands
Foundation for Basic Research of Matter; L. Moi from Siena University; A.Dalgamo and
G. Victor from the Harvard-Smithonian Center for Astrophysics for the opportunity to
work with their groups and to discuss the optical problems at seminars.

The work was partially supported by the International  Science Foundation,
the Russian Foundation for Basic Research, and the Krasnoyarsk Region
Science Foundation.

\end{document}